\documentclass{aa}
\usepackage{graphicx}
\usepackage{txfonts}
\newcounter{Rco}

\newcommand{\ea}{et al.\,}
\newcommand{\logg}{\mbox{$\log g$}}
\newcommand{\loggw}[1]{\mbox{$\log g\hspace{-0.5mm} =\hspace{-0.5mm}  #1$}}

\newcommand{\ab}[1]{\mbox{Fig.\,\ref{#1}}}
\newcommand{\sA}[1]{\mbox{(Fig.\,\ref{#1})}}

\newcommand{\Teff}{\mbox{$T_\mathrm{eff}$}}
\newcommand{\Teffw}[1]{\mbox{$\Teff\hspace{-0.5mm} =\hspace{-0.5mm} #1\,\mathrm{kK}$}}
\def\aspc#1#2{The ASP Conference Series Vol\@. #1 (San Francisco: ASP), p\@. #2}
\begin{document}
   \headnote{Research Note}
   \title
   {On the discovery of an enormous ionized halo \\ around the hot DO white dwarf \object{PG\,1034+001}
    }
 
   \author{T\@. Rauch$^{1, 2}$\thanks{Visiting Astronomer, German-Spanish Astronomical Center, Calar Alto, 
           operated by the Max-Planck-Institut f\"ur Astronomie Heidelberg jointly 
           with the Spanish National Commission for Astronomy} \and F\@. Kerber$^3$ \and E.-M\@. Pauli$^1$}
   \offprints{T\@. Rauch}
   \mail{Thomas.Rauch@sternwarte.uni-erlangen.de}
 
   \institute
    {Dr.-Remeis-Sternwarte, Sternwartstra\ss e 7, D-96049 Bamberg, Germany
    \and
     Institut f\"ur Astronomie und Astrophysik, Abteilung Astronomie, Sand 1, D-72076 T\"ubingen, Germany
    \and
    Space Telescope --- European Coordinating Facility, Karl-Schwarzschild-Stra\ss e 2, D-85748 Garching, Germany 
    }

     \date{Received December 2003 / Accepted 2004}

   \authorrunning{T\@. Rauch et al\@.}
   \titlerunning{On the Planetary Nebula around \object{PG\,1034+001})}
   \abstract{Very recently, the discovery of the {\it largest known planetary 
nebula on the sky} surrounding the DO white dwarf \object{PG\,1034+001} 
with an apparent diameter of about 2\degr, corresponding 
to a linear diameter of 3.5 - 7.0 pc at the likely distance of 100 - 200 pc, 
has been reported by Hewett \ea (2003). A careful inspection of available sky 
survey data has now shown that this planetary nebula, \object{Hewett\,1}, 
is surrounded by an elliptical emission shell with an apparent diameter of 
6\degr\,$\times$\,9\degr\  
($16.2^{+6.1}_{-4.5} \times 24.3^{+9.1}_{-6.8}\, \mathrm{pc}$ at
$d = 155^{+58}_{-43}\ \mathrm{pc}$).  
A further emission structure, detected northeast of
the central star may indicate another shell with a size of 10\degr\,$\times$\,16\degr. 
>From presently available observational data we do not have indications
whether the emission arises from material which was ejected from \object{PG\,1034+001}
or from ionized ambient ISM.

Improved proper motion data combined with radial velocity and 
distance from the literature have enabled us to derive a Galactic orbit for 
the central star \object{PG\,1034+001}. Its thin disk orbit and the 
morphology of the first halo suggest that the nebula is in an advanced stage 
of interaction with the interstellar medium. 
        }
   \maketitle

\section{Introduction} 
\label{int} 
\object{PG\,1034+001} ($\alpha_\mathrm{2000} = 10^\mathrm{h} 37^\mathrm{m} 04\fs0$,
                       $\delta_\mathrm{2000} = -0\degr 8\arcmin 20\arcsec$)
is a very hot, hydrogen-deficient DO white dwarf. 
Werner \ea (1995) determined \Teffw{100^{+15}_{-10}}, \loggw{7.5\pm 0.3} (cgs)
by means of NLTE model atmosphere techniques. From its position in the 
$\log g$ - $\log \Teff$ plane \sA{evo},
they concluded that \object{PG\,1034+001} is a
successor of the hydrogen-deficient PG\,1159 stars.

Very recently, Hewett \ea (2003) found by investigation of spectra from the Sloan Digital Sky Survey (SDSS,
York \ea 2000) that, at a distance of $155^{+58}_{-43}\ \mathrm{pc}$ (Werner \ea 1995),
a planetary nebula (PN) with a linear diameter of about 3.5 - 7.0\,pc surrounds \object{PG\,1034+001}.
This is larger than the PN \object{Sh\,2-216} with a size of 3.5\,pc at distance of $130^{+9}_{-8}\ \mathrm{pc}$
(Harris \ea 1997). Only one PN, namely \object{PN\,G080.3-10.4} (\object{MWP\,1})
surrounding the PG\,1159 star \object{RX\,J2117.1+3412} (Appleton \ea 1993), 
may be physically larger. At a distance of
$1.4^{+0.7}_{-0.5}\ \mathrm{kpc}$ (Motch \ea 1993), its apparent diameter 
of 14\arcmin 45\arcsec\ (Rauch 1997) corresponds to a physical size of 
$6^{+3}_{-2}\ \mathrm{pc}$.

However, whatever an improvement of the distance determinations for the PN around \object{PG\,1034+001}
or \object{RX\,J2117.1+3412} may yield --- our inspection of 
the Southern H-Alpha Sky Survey Atlas (SHASSA, Gaustad \ea 2001)
has shown two much more extended
emission structures surrounding \object{PG\,1034+001} which might make 
this object the largest PN known to date. In the following, we will describe 
our discovery in more detail.

\section{Examination of the SHASSA images}
\label{shassa}

A close inspection of SHASSA field \#214 \sA{sh214}, centered on 
$\alpha_\mathrm{2000} = 01^\mathrm{h} 43^\mathrm{m}$ and 
$\delta_\mathrm{2000} = -0\degr 14\arcmin$), 
shows in addition to the planetary nebula  discovered by Hewett \ea (2003)
two larger, elliptic structures which are apparently surrounding
\object{PG\,1034+001}. The inner has a diameter of 
6\degr\,$\times$\,9\degr\  corresponding to linear diameter of
$16.2^{+6.1}_{-4.5} \times 24.3^{+9.1}_{-6.8}\, \mathrm{pc}$ at
$d = 155^{+58}_{-43}\ \mathrm{pc}$.
The wider shell (only a fragment is visible in the NE,
see \ab{sh214}) has an estimated diameter of
10\degr\,$\times$\,16\degr. 
At a galactic latitude of +47{\degr}
filaments forming an elliptical shell are pretty rare in the SHASSA survey.
Therefore the likelyhood of a chance superposition with \object{PG\,1034+001}
is very small.

\begin{figure}[ht]
  \resizebox{\hsize}{!}{\includegraphics{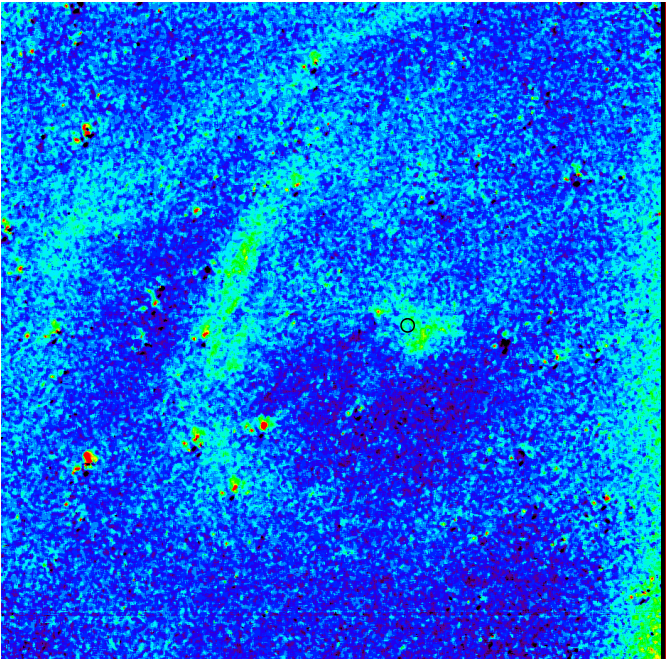}}
  \caption[]{SHASSA field \#214. \object{PG\,1034+001} is marked by a circle.
             East is left and North is up. 
             The field size is about 13\degr\,$\times$\,13\degr\ with a
             pixel size of 0\farcm8.
             Note the elliptic emission structure lying from SSE to NW, most
             prominent at NE side. The diameter is about 6\degr\,$\times$\,9\degr.
             A second, emission structure further outside can be seen in the NE.
            }
  \label{sh214}
\end{figure}

These structures are even larger than the ionized halo of ambient interstellar matter
(4\degr\,$\times$\,5\degr,  17\,$\times$\,21\,pc at a distance of 240\,pc)
around the PN \object{A\,36} which was recently discovered by McCullough \ea (2001) also using SHASSA.

\section{\object{PG\,1034+001} --- the exciting star?}
\label{exc}

Since the newly found emission structures are located at a significant 
distance (up to about 12\,pc)
from \object{PG\,1034+001} the question arises whether this object can actually
excite
them. Werner \ea (1995) proposed that \object{PG\,1034+001} is a descendant
from the PG\,1159 stars. By comparing the position of \object{PG\,1034+001}
with theoretical evolutionary tracks of born-again post-AGB stars 
we can determine a stellar mass of $M = 0.62 \pm 0.06 \mathrm{M_\odot}$ \sA{evo}
which is slightly higher than the one found by Werner \ea (1995, 
$M = 0.59^{+0.14}_{-0.03}\mathrm{M_\odot}$) who compared with 
evolutionary tracks of Wood \& Faulkner (1986) for He burning post-AGB stars
which did not experience a final helium-shell flash. 

>From $M = 0.62 \pm 0.06\mathrm{M_\odot}$, we can calculate a stellar radius
$R = 0.0232^{+0.0112}_{-0.0076} \mathrm{R_\odot}$ for \object{PG\,1034+001}.
The numbers of emergent photons with energies higher than 13.6\,eV
 and the corresponding Str\"omgren radii 
$R_{\mathrm{H}\,\mathsc{ii}}$
are summarized in Table\,\ref{rhii}.

\begin{table}
\caption{Str\"omgren radii for \object{PG\,1034+001}. The pure He models are calculated with
         \loggw{7.5}, $R_\mathrm{H\,II}$ is calculated with an electron temperature
         $T_\mathrm{e} = 10000 \mathrm{K}$, an electron density 
         $n_\mathrm{e} = 3 / \mathrm{cm^3}$, and a volume filling factor of 
         $\epsilon = 0.25$, and
         $R = 0.0232 \mathrm{R_\odot}$.}
\label{rhii}
\begin{tabular}{ccr@{.}l}
\hline
\hline
\noalign{\smallskip}
\Teff / kK & $N_{13.6}$ / cm$^2$ / sec & \multicolumn{2}{c}{$R_\mathrm{H\,II}$ / pc} \\
\hline       
\noalign{\smallskip}
 90 & $1.529\cdot 10^{26}$ &  6&5 \\
100 & $2.076\cdot 10^{26}$ &  7&2 \\
120 & $3.384\cdot 10^{26}$ &  8&5 \\
140 & $5.088\cdot 10^{26}$ &  9&8 \\
160 & $7.217\cdot 10^{26}$ & 11&0 \\
\hline
\end{tabular}
\end{table}

Werner \ea (1995) determined a current \Teffw{100^{+15}_{-10}}, but of course 
\object{PG\,1034+001} has
passed through a much hotter stage (\Teffw{150}) during its evolution \sA{evo}, only about 30\,000 years ago
(Bl\"ocker 1995).
Since the recombination time scale in such a low-density nebula is longer than 30\,000 years,
\object{PG\,1034+001} is most likely the exciting star of the nebula and the halo.

\begin{figure}[ht]
  \resizebox{\hsize}{!}{\includegraphics{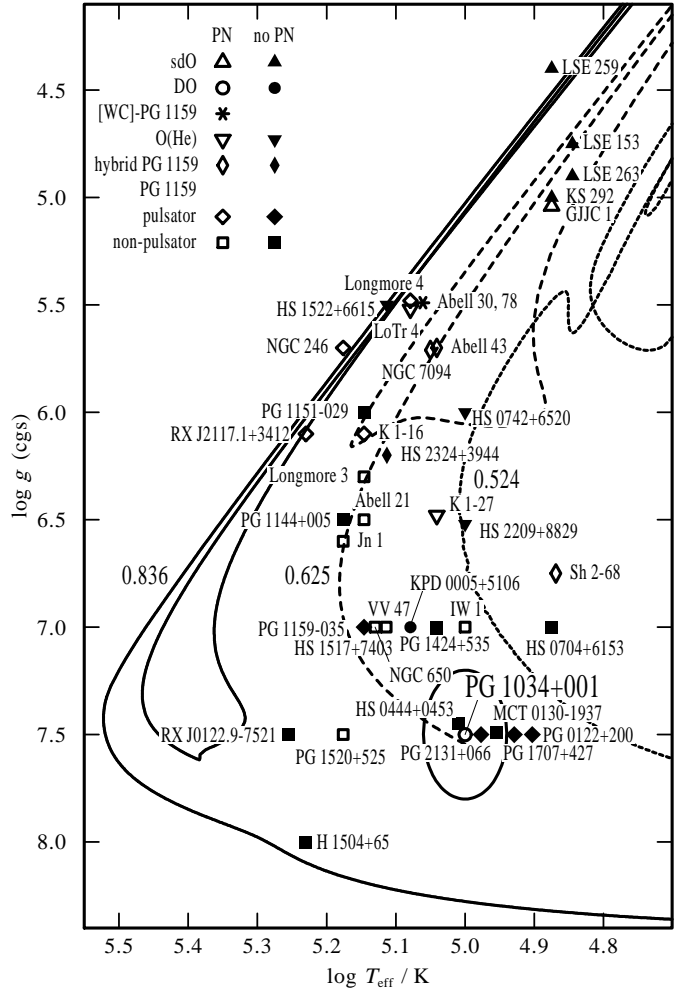}}
  \caption[]{Position of \object{PG\,1034+001} and related objects in the $\log \Teff - \logg$ plane 
             compared with evolutionary tracks (labeled with the respective stellar masses in M$_\odot$) 
             of born-again post-AGB stars (Bl\"ocker 1995). The ellipse shows the error range
             in the spectral analysis of Werner \ea (1995).
            }
  \label{evo}
\end{figure}

\section{Galactic orbit of \object{PG\,1034+001}}
\label{orbit}

\begin{figure}[ht]
  \resizebox{\hsize}{!}{\includegraphics{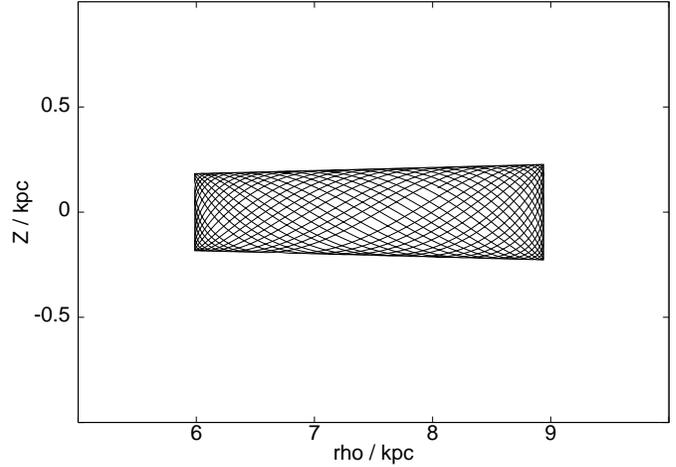}}
  \caption[]{Galactic orbit of \object{PG\,1034+001}.
             $\rho$ ist the distance from the Galactic Center, $z$ is the distance perpendicular to the disk.
            }
  \label{orb}
\end{figure}

Proper motion (PM) of \object{PG\,1034+001} had already been reported by Hewett 
\ea (2003). We have found two additional independent measurements in the 
UCAC2 and USNO-B catalogues, which confirm and improve upon the previous 
value (Tab.\,\ref{tab_pm}). For the USNO-B entry errors smaller than 
4\,mas\,yr$^{-1}$ are considered too optimistic (Gould 2003), and have been 
set to 4\,mas\,yr$^{-1}$.
This puts us in a position to calculate
a Galactic orbit for the star, adopting a distance of $155^{+58}_{-43}\ \mathrm{pc}$ 
as derived by Werner \ea (1995). 
In terms of radial velocity one can improve upon the previously reported value by
using the velocity of the photospheric absorption lines found in the IUE 
spectra (Holberg \ea 1998) which result in a value of 
$50.83\,\pm\,1.33 \mathrm{km\,s^{-1}}$.
An orbit representative of the actual motion of the star is then derived by 
integrating back in time, with a code from Odenkirchen \& Brosche (1992), 
using a Galactic potential 
from Allen \& Santillan (1991). 
It should be noted that this orbit is an idealized one as 
the Galactic potential used is simplified and scattering processes 
between individual stars are neglected. 
Nevertheless orbital parameters (such as the eccentricity) 
can be computed and combined with velocities to gain 
information on population membership. 
We use the classification scheme presented in Pauli \ea (2003), 
where additionally to the position in the classical $U$-$V$-diagram, 
the position in the angular-momentum-eccentricity diagram and 
the Galactic orbit are considered. 
We find that all three criteria favor a thin disk membership 
of \object{PG\,1034+001}. 
The orbit found (Fig.\,\ref{orb}) is typical for a thin disk star  
showing both low inclination and eccentricity.
We use the Galactic velocity reference system in which the local
standard of rest has (U,V,W) = (0,220,0) km/s.
U is in radial direction towards the galactic center, V in the direction
of the galactic rotation and W in the direction perpendicular to the
Galactic disk. 
Details of the procedure can be found in Pauli \ea (2003).

It is interesting to note that a number of stars with rather similar PM values
are present in the vicinity. Within a search radius of 2\degr\ we found 15 
stars which have the same PM within 3\,$\sigma$. While small number statistics 
prevent us from drawing any firm conclusions this indicates that 
\object{PG\,1034+001} might be part of a group of stars with a common origin.
Most of these stars are faint but one object has a magnitude of about 11, 
which will make it easy to obtain high quality spectra and investigate its 
properties. Using stars within a radius of 1\degr, we have also investigated 
the reddening as a function of distance towards \object{PG\,1034+001}. We 
find the distance compatible with the range of 100 to 200\,pc reported 
before, with a possible preference for the lower end of the range.

\begin{table}
\caption{Proper motion (in mas\,yr$^{-1}$) for \object{PG\,1034+001} from different sources.}
\label{tab_pm}
\begin{tabular}{cccc}
\hline
\noalign{\smallskip}
& Hewett \ea (2003) & UCAC2 & USNO-B\\
\hline       
\noalign{\smallskip}
$\mu(\alpha$)cos($\delta)$ &  	$-86 \pm 5$ & $-83.7 \pm 3.3$ & $-80 \pm 4$\\
$\mu(\delta$)& $+31 \pm 5$ & $+28.9 \pm 1.5$ & $+32 \pm 4$\\
\hline
\end{tabular}
\end{table}

\section{Interaction with the ISM}
For highly evolved PNe the density in the shell can become so low that the ram 
pressure of the moving PN becomes comparable to the thermal pressure of the 
surrounding interstellar medium (ISM). At this point it will start to 
interact with the ISM, a process described in detail by Borkowski \ea (1990)
and Soker \ea (1991). The examples for interaction with the ISM
range from mild cases like Sh\,2-216 (Tweedy \ea 1995) to the extreme case 
of the PN abandoned by its CS, Sh\,2-174 (Tweedy \& Napiwotzki 1994). 
Many more examples of PNe interacting with the ISM have been reported recently,
(Tweedy \& Kwitter 1996, Xilouris et al 1996, Kerber \ea 2000,
Rauch \ea 2000).\newline
Kerber \ea (2004) have most recently demonstrated that the
orbital motion of the PN is indeed the root cause for this interaction with the 
ISM. Moreover, they showed that the parameters of the orbit, most notably 
eccentricity and inclination, both of which can have a pronounced effect on the velocity
relative to the ISM, strongly influence the kind and degree of interaction.
Fig.\,\ref{sh214} shows the 7\degr\ halo as an incomplete ellipse that is well
defined in the E and fades in the N and SW. Unfortunately the W edge of the 
halo is located close to border of the frame in the SHASSA survey and flat 
fielding is critically for analysis of such a low surface brightness structure.
The adjacent field 213 shows diffuse emission which lacks any arclike 
structure and which is about a factor 1.5 fainter than the arc in the E. 
This diffuse emission could be part of a varying background.
It seems, therefore that no well defined halo exists in the W. The major axis 
of the halo has a position angle (PA) of about 310\,$\pm$\,15\degr\ roughly 
consistent  with the PA of the proper motion of 289 $\pm$ 2\degr.
This structure can be understood in terms of an advanced interaction with the 
ISM.
In this scenario the leading edge of the halo has already been destroyed and 
its matter has mixed with the ISM, 
forming the faint diffuse emission in the W, 
while the trailing eastern edge, is still 
rather well defined. Since \object{PG\,1034+001} is located close to the 
Galactic plane it is likely to reside in relatively dense ISM and interaction 
is possible even at modest relative velocities. The thin disk orbit found for 
\object{PG\,1034+001} also suggests a mild form of interaction, which because 
of the long time scales involved can be expected to be in an advanced stage.

For the outer halo a similar 
line of reasoning is possible but remains highly speculative because of the limited
information we have on this structure.
The inner nebula, \object{Hewett\,1}, is highly filamentary and the CS is 
not in the geometric center of the nebula. As already pointed out by Hewett 
\ea (2003) the spatially varying ratios of [\ion{O}{iii}], H$\alpha$, and 
[\ion{N}{ii}] can also be explained by an interaction with the ISM. 
Since the leading edge of the halo is missing it is also likely that the ISM 
can interact with the inner nebula. 
In the case of \object{PG\,1034+001} particular caution is due with this 
interpretation, since it is a DO white dwarf, and it probably has 
experienced a final helium-shell flash in its past, which produces a second 
PN with a very different, hydrogen-deficient, chemical composition. Also, 
this second PN can 
be ejected in a highly non-spherically symmetric manner, as seen in 
\object{A\,30} (Borkowski \ea 1995) and  \object{V4334\,Sgr} 
(Kerber \ea 2002a).

\section{Conclusions}
\label{con}
   
We have discovered a huge elliptical halo with a linear diameter of 
$16.2^{+6.1}_{-4.5} \times 24.3^{+9.1}_{-6.8}\, \mathrm{pc}$ at
$d = 155^{+58}_{-43}\ \mathrm{pc}$
surrounding the PN \object{Hewett\,1} and its exciting star \object{PG\,1034+001}
on SHASSA images.
An even larger emission structure is also visible but much fainter.
>From presently available observational data we do not have indications
whether the emission arises from material which was ejected from \object{PG\,1034+001}
or from ionized ambient ISM.
Clearly better spectra and images as well as a better distance determination 
are needed in order to understand the nature of the PN surrounding \object{PG\,1034+001}.

It is worthwhile to note that the largest PNe known so far,
\object{Sh 2-68} (Kerber \ea 2002b), 
\object{Hewett\,1}, and
\object{MWP\,1},
have been found around so-called born-again post-AGB type central stars
(Iben \ea 1983). Since their exciting stars have quite different masses
(0.50\,M$_\odot$, 0.62\,M$_\odot$, and 0.83\,M$_\odot$, respectively, see \ab{evo}),
the size of the PN does not appear to be related to the final mass of the
ejecting star.

We have derived the Galactic orbit of \object{PG\,1034+001} and determined 
that it belongs to the thin disk population. The kinematical parameters and 
the morphology of the halo indicate that the nebula is in an advanced stage 
of interaction with the ISM. Since this is one of the closest and 
most evolved PNe, it is ideally suited for detailed study of this process 
which governs the return of processed matter to the ISM.

\begin{acknowledgements}
This research was supported by the DLR under grant 50\,OR\,0201 (TR) and by the
DFG under grant Na\,365/2-1 (EMP). 
EMP also wishes to express gratitude to the Studienstiftung des Deutschen Volkes for a grant.
This research has made use of the SIMBAD Astronomical Database, operated at CDS, Strasbourg, France.
\end{acknowledgements}

\end{document}